\documentclass[12pt]{article}
\usepackage{amsmath}
\usepackage{graphicx}
\usepackage{psfrag}
\usepackage{epsf}
\usepackage{algorithmic}
\usepackage{algorithm}
\usepackage{enumerate}
\usepackage{natbib}
\usepackage{float}

\newcommand{\blind}{0}

\begin{document}

\def\spacingset#1{\renewcommand{\baselinestretch}%
{#1}\small\normalsize} \spacingset{1}

%%%%%%%%%%%%%%%%%%%%%%%%%%%%%%%%%%%%%%%%%%%%%%%%%%%%%%%%%%%%%%%%%%%%%%%%%%%%%%

\if0\blind
{
 \title{\bf Orbital Separation Amplification in Fragile Binaries with Evolved Components}
  \author{ K.B. Johnston and T. D. Oswalt\footnote{Address: Dept. of Physics \& Astronomy, Florida Institute of Technology, 150 West University Blvd., Melbourne, FL 32901; e-mail: kjohnst2000@fit.edu and toswalt@fit.edu} \\ D.Valls-Gabaud\footnote{Address: LERMA, CNRS UMR 8112, Observatoire de Paris, 61 Avenue de l'Observatoire, 75014 Paris, France.; e-mail: David.Valls-Gabaud@obspm.fr}}
  \maketitle
} \fi

\if1\blind
{
  \bigskip
  \bigskip
  \bigskip
  \begin{center}
    {\LARGE\bf Title}
\end{center}
  \medskip
} \fi

\bigskip
\begin{abstract}
The secular stellar mass-loss causes an amplification of the orbital separation in fragile, common proper motion, binary systems with separations of the order of 1000 A.U. In these systems, companions evolve as two independent coeval stars as they experience negligible mutual tidal interactions or mass transfer. We present models for how post-main sequence mass-loss statistically distorts the frequency distribution of separations in fragile binaries. These models demonstrate the expected increase in orbital seapration resulting from stellar mass-loss, as well as a perturbation of associated orbital parameters. Comparisons between our models and observations resulting from the Luyten survey of wide visual binaries, specifically those containing MS and white-dwarf pairs, demonstrate a good agreement between the calculated and the observed angular separation distribution functions.
\end{abstract}

\noindent%
{\it Keywords:}  binaries: general - stars: statistics - white dwarfs

\newpage
\spacingset{1.45} % DON'T change the spacing!
\section{Introduction}

Mass-loss due to various processes during the lifetime of a star will increase the semi-major axis of a wide binary \citep{DOM:63, VG:88}, what we refer to here as orbital separation amplification. It is expected that mass-loss during the stellar lifetime, in particular Post-Main Sequence (PMS) mass-loss, will statistically distort a frequency distribution of weakly interacting wide binary (Fragile Binary, FB) separations. Visual confirmation of such processes would provide tests of current theories of stellar evolution and fundamental information on the dynamics of Galactic structures \citep{W:88}. 

As discussed by \cite{OS:07} and \cite{CHAN:07}, FB's provide insight into several astrophysically important problems that cannot be easily addressed by other means. In particular, such fragile pairs provide valuable constraints on the Galactic tidal potential, stellar and ISM perturbations, and PMS mass-loss, all of which will affect orbital separations. A number of early papers have addressed this topic: \cite{EGG:65} studied the colors and magnitudes of 228 visual binaries (VBs), deriving an $M_v$ vs. $R-I$ calibration from trigonometric parallaxes. \cite{VD:80} showed that, using proper motions from micrometer measurements, a pair could be deemed gravitationally bound and thus, through statistical studies and interferometry, orbital parameters could be determined. \cite{TRIM:74} performed a statistical study of the observed populations of binary systems (798 VBs) and was one of the first to note the peak in the mass ratio (q) distribution at $q\sim1$. 

In recent years, computing power has improved to the point that modeling the evolution of whole stellar binary populations is now possible. Examples include: \cite{W:88, VG:88}, and more recently \cite{HTAP:01}. Prior work has shown evidence for orbital separation amplification by as much as half an order of magnitude in binaries, where at least one of the partners has undergone PMS mass loss \citep{HAD:66, G:86, WO:92}. Considering that the majority of all stars are born with one or more companions as a result of the initial conditions required for a stable system \citep{M:94}, the importance of understanding how mass-loss affects the evolution of binary systems cannot be over stated. In addition, the initial-to-final mass relation (IFMR) for WD stars is at present very poorly constrained \citep{WEID:93, ZH:11}. Proper modeling of FB orbital separation amplification can provide a valuable independent estimate of this important relation.

The distribution of $WD+MS$ pair separations is most strongly affected by the PMS (i.e. red giant phase) mass-loss. FB systems typically have component separations of $\sim10^3 AU$ \citep{ZI:84}; this results in negligible tidal interactions and mass transfer between companions. Such pairs evolve as if they are two separate but coeval stars. Once mass-loss has ended, orbital evolution is limited to the effects of external forces such as stellar encounters, giant molecular cloud encounters, gravitational tidal forces, and rare encounters with other stars \citep{CHAN:04,JIAN:10} .

In this study, which updates and improves upon our previous work \citep{JOHN:07}, we compare a theoretically generated population of FB angular separations to the observed angular separation distributions of FBs found in a large-scale survey. We constructed a simulation code, which we refer to here as the Double Star (DS) program, to generate the initial parameters and distributions of the tested population. The DS program evolves the binary components using the Single Star Evolution (SSE) code \citep{HPT:00} over the lifetime of the simulation. The DS program is also responsible for the evolution of individual orbital elements and translation of the modeled physical properties into observable features. The results of the translation is a population of theoretical FBs, with observable properties.

\section{Observational Data}

It should be possible to produce a statistically significant frequency distribution of angular separations for FBs with various constituents, by sampling a large survey such as the Luyten Double Star (LDS) Catalog \citep{LDS:69}. While the LDS catalog, with 6124 pairs and multiples, does not have the volume of more recent surveys such as the Sloan Digital Sky Survey (SDSS), it provides a well-observed large sample to compare to our theoretical models. The identification of FB candidates requires selection of an upper bound separation. Luyten found pairs separated by as much as $400"$. Due to the POSS survey plate scale, Luyten's minimum angular separation limit was $\sim0.5"$. We will use this sample as a first test of our programme, while we note that further samples of fragile binaries are currently being constructed (e.g., \citealt{SES:08, QUIN:09, LONG:10, ZHA:11}).

\section{Initial Parameters}

To build a synthetic model of the local Galactic FB population, we first proposed a set of initial parameters that defined the individual components of our binaries, as well as the binary system separations. We used a combination of published functions from previous theoretical models and observational surveys. When there were no such references, or when the functions published were not appropriate for FBs, we implemented physically reasonable distributions.

\subsection{Mass Generating Function}
Following \cite{KTG:93} we assumed a Salpeter distribution of ZAMS stars using a variation of Eggleton's mass generating function (Equation \ref{eq:a1}).

\begin{equation}\label{eq:a1}
M(x) = 0.33( \frac{1.0} {(1.0 - x)^{0.75} + 0.04(1.0 - x)^{0.25}} - \frac{1.0}{1.04} (1.0 - x)^2 )
\end{equation}

\begin{figure}[h!]
  \centering
    \includegraphics[scale=1.0]{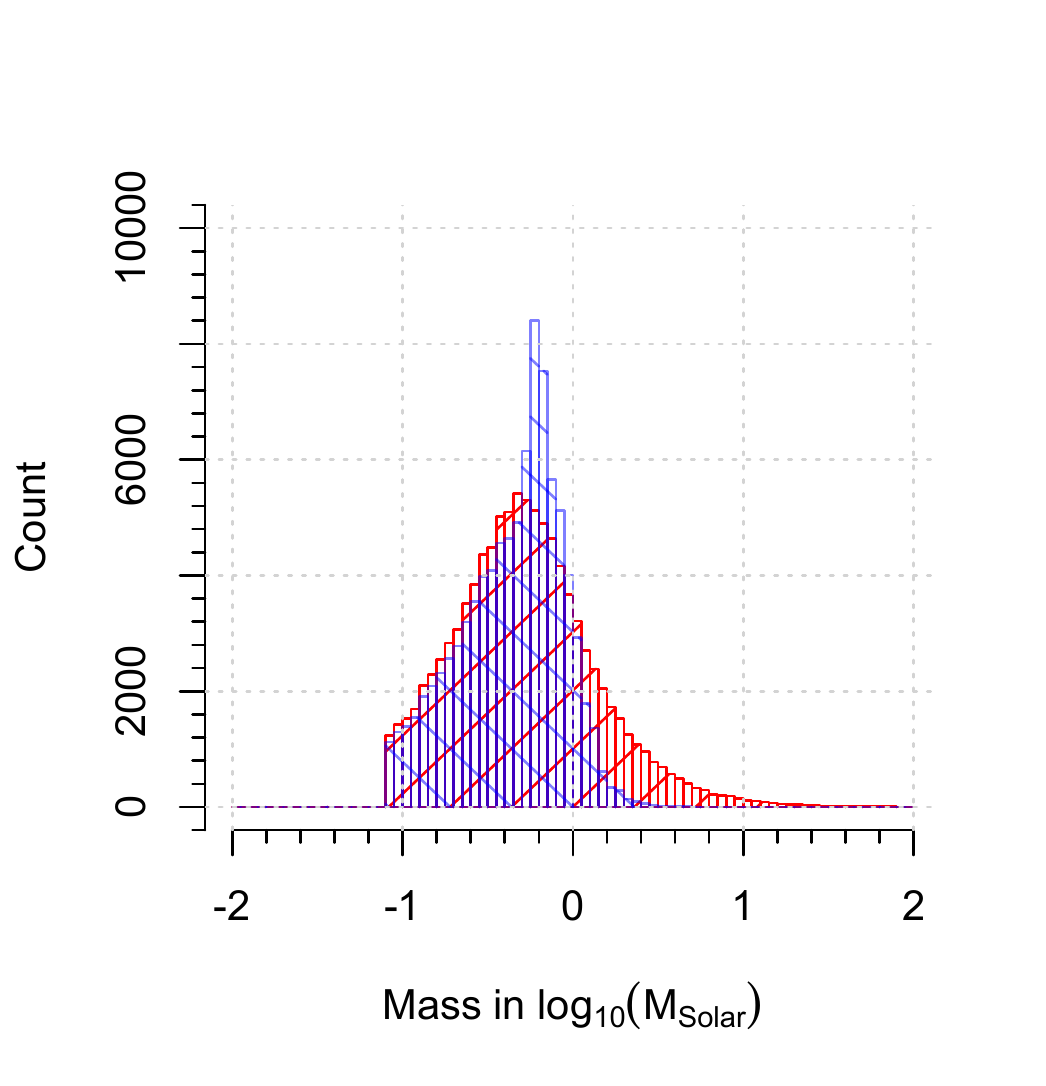}
 \caption{Histogram of the distribution of primary masses from initial generation (in red with 45 deg. hatching) and after evolution and orbital expansion (in blue with -45 deg. hatching). The difference in population size is resulting from the dissolution of some binaries from the original population. Each bin is of size $0.05$ in $\log{(M_\odot)}$}\label{fig:a}
\end{figure}

Where x is a uniform randomly distributed number and M(x) is the randomly generated, restricted, mass of a ZAMS star.  For this project, we initially assumed a minimum mass of $0.07M_{\odot}$(spectral type-T), given by \cite{CPH:03}. From this we produced an initial distribution of primary masses (Figure \ref{fig:a}).

\subsection{Mass Ratio}

The mass ratio, $q$, is one of the most poorly determined variables in a wide binary system. Current observed distributions fall into two categories: (1) a distribution that peaks at $q\sim1$ \citep{TRIM:87} or (2) a bimodal distribution with peaks at $q\sim 0.2$ and $q\sim 0.7$ \citep{HALB:03}. The former are explained in terms of turbulence and fragmentation during the binary formation process \citep{LUC:79}. This tends to result in near equal clumps of would-be stars forming out of a single GMC. Bimodal distributions seem to be created by selection effects (spectroscopic methods vs. common proper motion) and evolutionary effects (binary capture, disruption, mass-loss, mass-transfer,etc.), for references of either see \cite{KRU:67, P:71, SHU:81}. For the purpose of finding an initial mass ratio function (IMRF), we will tend to use the discussion of distributions which tend towards a maximum of one in the set 0$\leq q \leq$1(unity-like), as these will be the most likely candidates to explain a mass-ratio distribution created by a GMC to stellar binary creation process. After multiple trials with various models, we chose a random uniform distribution, where 0$\leq q \leq$1 and $\lambda$ is a constant (Equation \ref{eq:b1}).

\begin{equation}\label{eq:b1}
f(q) = \lambda
\end{equation}

\begin{figure}[h!]
  \centering
    \includegraphics[scale=1.0]{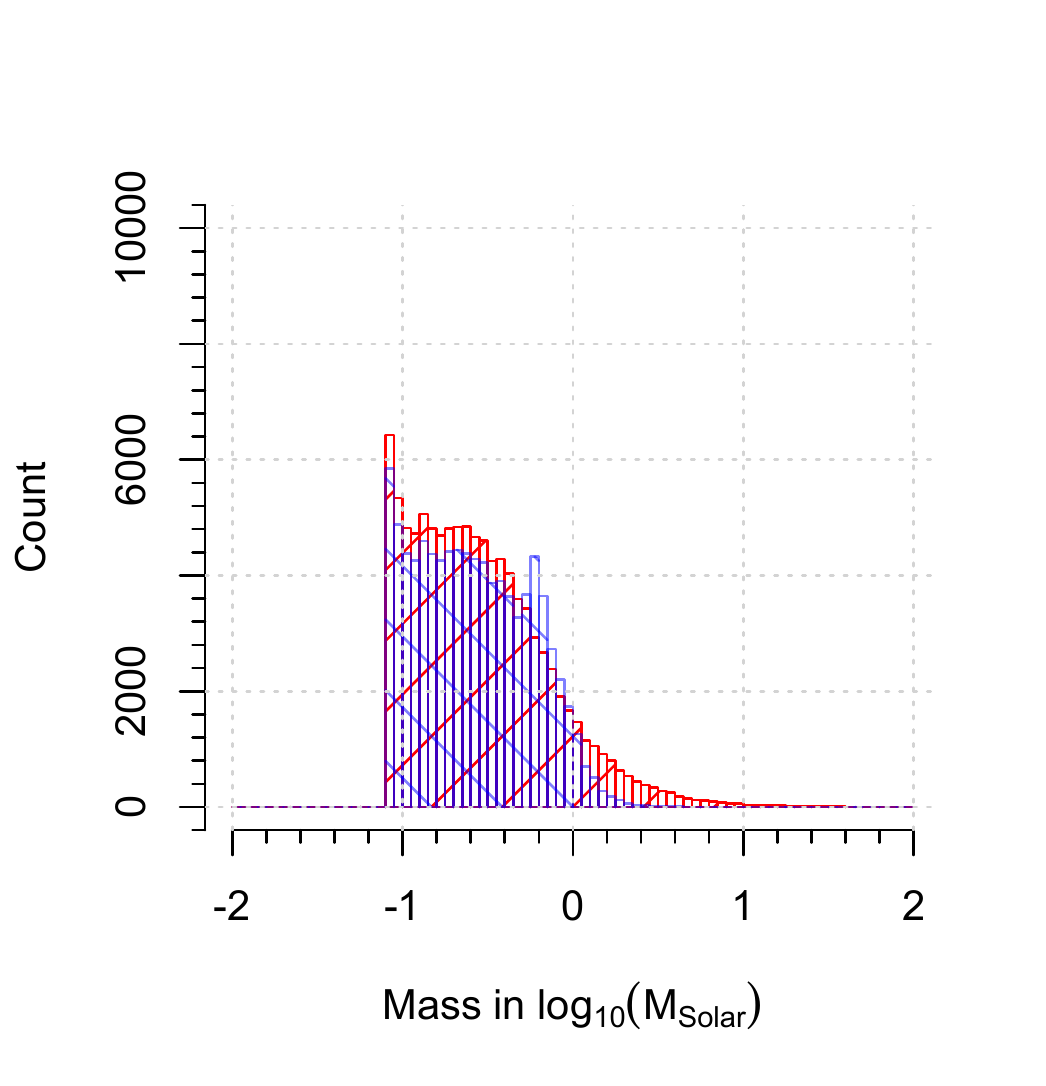}
 \caption{Histogram of the distribution of secondary masses from initial generation (in red with 45 deg. hatching) and after evolution and orbital expansion (in blue with -45 deg. hatching). The difference in population size resulted from the dissolution of some binaries from the original population.Each bin is of size $0.05$ in $\log{(M_\odot)}$}\label{fig:b}
\end{figure}

\begin{figure}[h!]
  \centering
    \includegraphics[scale=1.0]{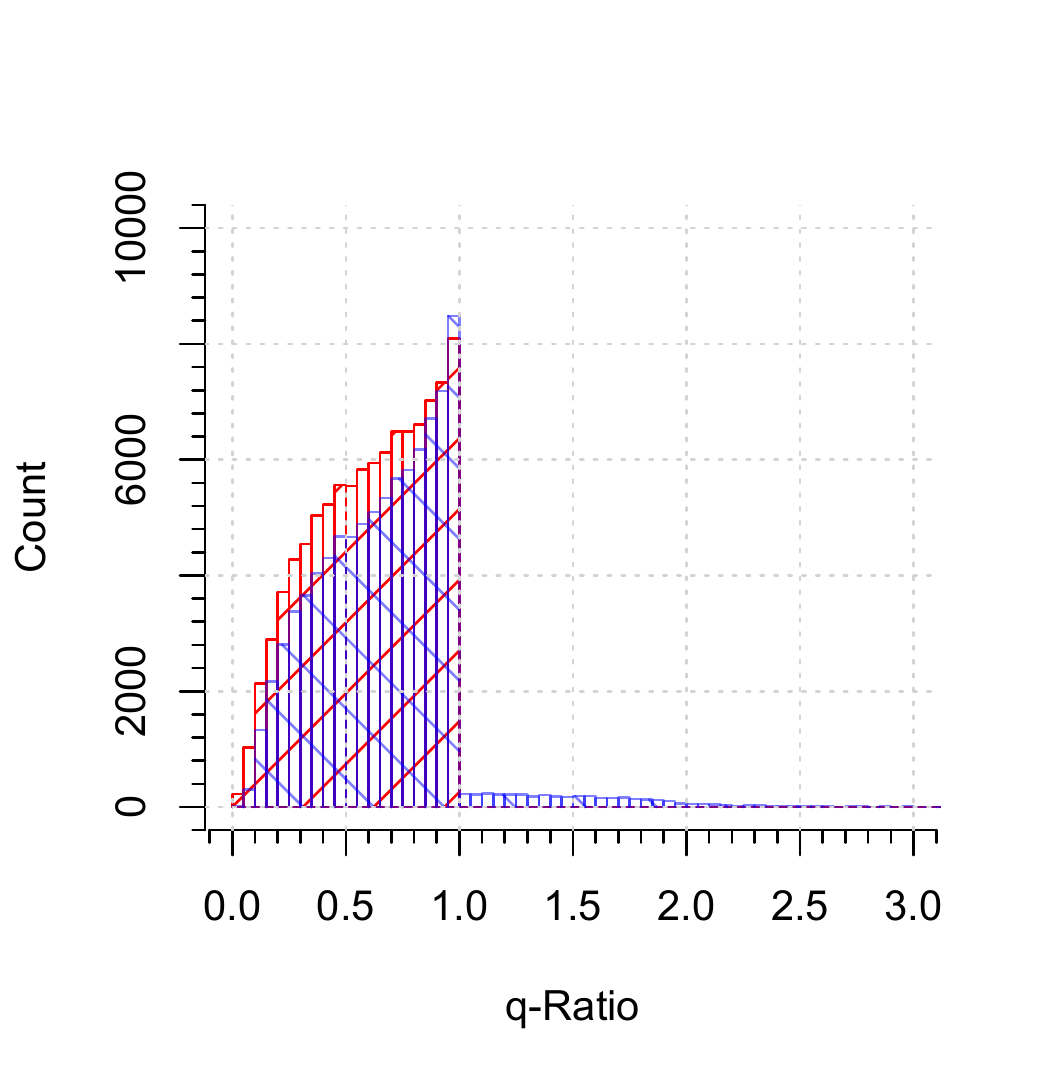}
 \caption{Histogram of the distribution of mass ratios from initial generation (in red with 45 deg. hatching) and after evolution and orbital expansion (in blue with -45 deg. hatching). The difference in population size resulted from the dissolution of some binaries from the original population. Bin size is of size $0.05$ in $q$-ratio}\label{fig:c}
\end{figure}

A constant mass ratio distribution function has the added advantage of producing an undistorted secondary mass distribution, implying that the same physical processes produce the primary and secondary components. Using an acceptance-rejection method of random variate generation \citep{LAW:05}, we produced the distribution of secondary masses seen in Figure \ref{fig:b}. In addition we restricted all lower mass components to $M_{star}\geq0.07M_{\odot}$. This has the added effect of producing a tendency for $q\sim1$, as seen in Figure \ref{fig:c}, in accord with prior observational studies.

\subsection{Orbital Parameter Distribution Functions}
In this paper, we focused on the development of two specific orbital element distributions, specifically the distribution of semi-major axis (separation) and the distribution of eccentricity. From these parameters, we develop elliptical binary mechanics for each binary in our initial population. The statistical characteristics of the separation distribution for the un-evolved binaries, as well as the distribution characteristics for the primary masses, secondary masses, and periods is given in Table \ref{tab:a} at the end of this section.

\subsubsection{Initial Separation Function}
The separations of binary systems is a difficult observable parameter, requiring distance, inclination, eccentricity and mass. One of the first and most often used references to separation distribution was given by \cite{O:24}, and implemented by \cite{ZI:84} and \cite{P:88}.  It is a simple uniform log distribution:

\begin{equation}\label{eq:c1}
f(a) = \log(a)
\end{equation}

Where $a \geq 100 A.U.$, approximately the closest separation where post-MS mass exchanges are unlikely to have occured. We transform equation \ref{eq:c1} into a random variate separation generating function (Equation \ref{eq:d1}).

\begin{equation}\label{eq:d1}
a(X) = 10^{2.0X+2.0}
\end{equation}

Where $0 \leq X \leq 1$ and $X$ is a uniformly distributed random number. Similarly, we used the separation and masses now assigned to each pair, along with Kepler's $3^{rd}$ law to calculate the period for each binary.

\begin{figure}[h!]
  \centering
    \includegraphics[scale=1.0]{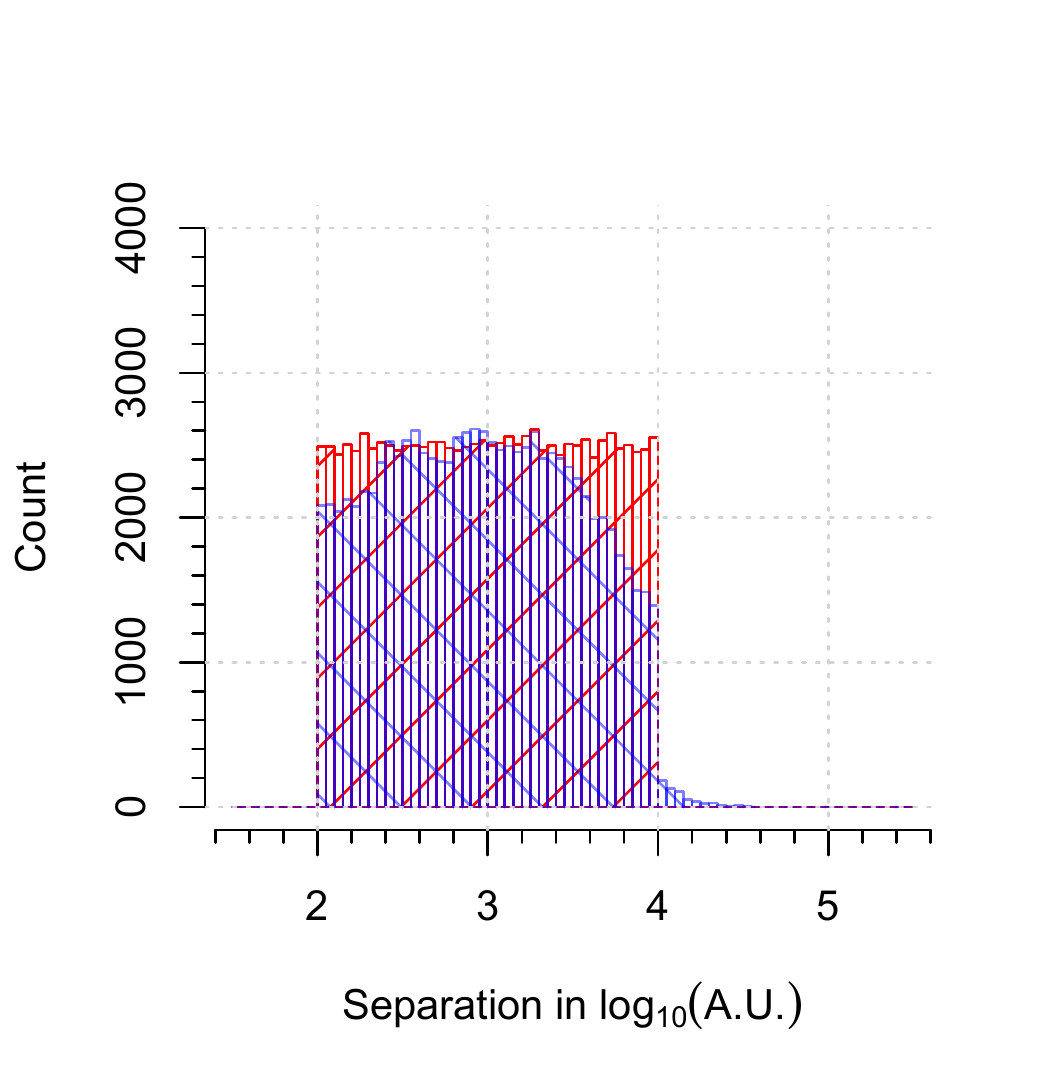}
 \caption{Histogram of the distribution of separation from initial generation (in red with 45 deg. hatching) and after evolution and orbital expansion (in blue with -45 deg. hatching). The difference in population size resulted from the dissolution of some binaries from the original population. Each bin is of size 0.05 in $\log{(A.U.)}$}\label{fig:d}
\end{figure}

\begin{figure}[h!]
  \centering
    \includegraphics[scale=1.0]{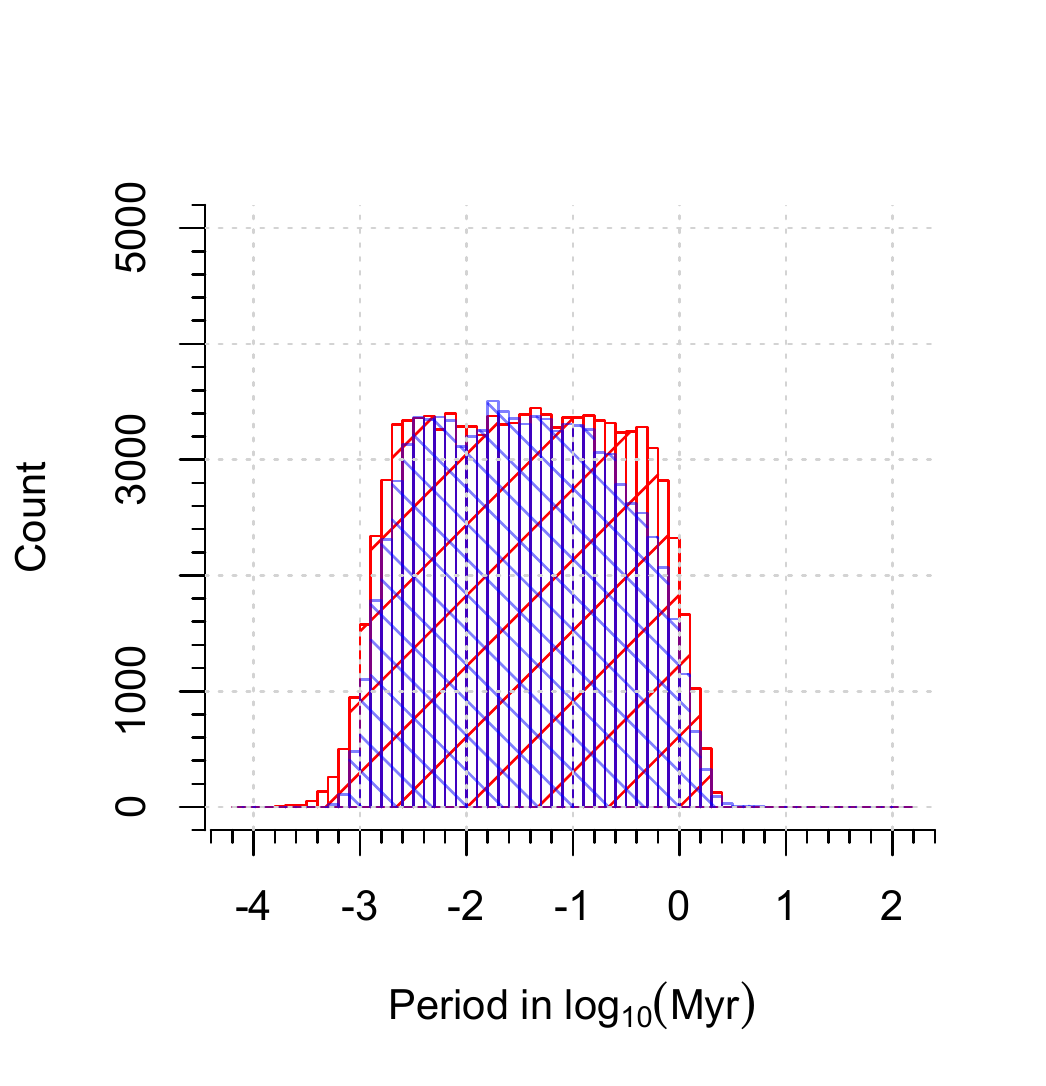}
 \caption{Histogram of the distribution of periods from initial generation (in red with 45 deg. hatching) and after evolution and orbital expansion (in blue with -45 deg. hatching). The difference in population size resulted from the dissolution of some binaries from the original population. Each bin is of size 0.1 in $\log{(Myr)}$}\label{fig:e}
\end{figure}

In addition to the physical limit of $a\geq100 A.U.$ for non-interacting pairs, we also imposed observed limitations on our binary pairs. We rely on the \cite{ZI:84} definition of CPMB/FB, VB, and SBs and limit the initial smallest separation that a common proper motion binary can have as $100 A.U.$. This is wide enough that the two components of the pair are individually resolvable and the effects of Roche Lobe interaction and component-to-component tidal interactions are minimal compared to Galactic effects and stellar GMC encounter perturbations. 

We additionally imposed a maximum of $10^4 A.U.$ , which appears to be the maximum theoretical separation in wide pairs \citep{WW:87} imposed by gravitational disruptions of the Galaxy and stellar perturbations. An algorithm was developed to impose this upper limit continuously over the evolution time frame and will be discussed below in section 4.3.

\subsubsection{Eccentricity}
The simulation derives the initial distribution of eccentricities following \cite{HEG:75}. The initial distribution of eccentricity for wide binary pairs is given in Equation \ref{eq:e1}.

\begin{equation}\label{eq:e1}
f(e) = 2e
\end{equation}

We transform this distribution into the random variate generating function (Equation \ref{eq:f1}).

\begin{equation}\label{eq:f1}
e(X) = \sqrt{X}
\end{equation}

Where $0 \leq e < 1$ and $X$ is $0 \leq X \leq 1$, a uniformly distributed random number. This distribution was derived based on assumptions of a Maxwellian particle distribution. It becomes apparent from the generating function that binaries, with this eccentricity distribution, are not favored to have initial circular orbits. In fact, using this distribution one can easily see that the expected initial density of circular orbits is zero. For wide binary pairs we can convince ourselves that this is true, as these pairs tend to have coeval components but have had little or no interaction beyond the gravitational binding which makes them a pair (as discussed in the mass ratio section). This creation via loose interaction would tend to result in more initially eccentric orbits.

\begin{table}[h!]
\caption{Statistical Characteristics of the Un-evolved Distributions}\label{tab:a}
\begin{center}
\begin{tabular}{c|cccccc}
\multicolumn{7}{c}{} \\
\hline
Metric & Average & StdDev & Skew & Kurtosis & Min & Max \\
\hline
Mass 1($\log{M_{\odot}}$)&	-0.28&	0.41&	0.58&	0.79&	-1.10&	1.93\\
Mass 2($\log{M_{\odot}}$)  &	-0.55&	0.40&	0.92&	1.03&	-1.10&	1.84\\
Separation ($\log{A.U.}$) &	3.00&	0.58&	-0.002&	-1.20&	2.00&	4.00\\
Period ($\log{Myr}$) &	-1.46&	0.826&	-0.008&	-1.08&	-4.04&	0.39\\
\hline
\end{tabular}
\end{center}
\end{table}

\subsection{Birth Rate}

Our experiments incorporated an initial stellar burst of $10^2$ binaries with \cite{WK:04} constant repeating burst model of the stellar birth rate, $10^2$ binaries being created every $10 Myrs$, resulting in a total initial population of 100,100 binaries. All bursts have identical initial factors and distributions, resulting in a continuous and smooth distribution \citep{HPT:00} in the initial population. The only references that challenge this approach, suggest that metallicity may play a role in distorting the initial distribution function. However, the region of space that we are attempting to model is limited to the solar neighborhood (i.e. distances under $100 pc$), FBs in this region tend to have solar metallicitys. We therefore assumed solar metallicity for all ZAMS stars. Future work may address a more realistic history of star formation rate, dervied for the solar neighbourhood \citep{HER:00}.

\subsection{Color Estimation}

The goal of this project is to compute a theoretical population of FBs that accurately compared to observed angular separation distributions. This requires a transformation of our physical characteristics for each binary system into observables such as color, magnitude and angular separation. We employed a mix of \cite{KV:06, K:79, HAW:02} temperature vs. color relations for MS and WD stars, to produce the $ugriz(J)$ absolute magnitudes required for comparison. 

We divided stars into four categories: MLT, WD, MS, and other. MLT refers to the spectral type of the star: M, L or T, corresponding to effective temperatures from $\sim3800K$ to $1000K$ \citep{R:00, R:02}. The spectral type of a star below $3400K$ is assumed to be dependent only on temperature. We justify this estimation with three points: (1) Reliable statistics are not available for surface gravities of low mass stars (early-T and late-L); (2) The current classification scheme for MLT dwarfs is based mostly on line features. A number of flux ratios ($H_2O, CO_2, CH_4,$etc) show linear relations with spectral type \citep{BURG:02}; (3) \cite{STE:01} used a hybrid approach to show that temperatures and spectral types scale linearly with each other for L-type dwarfs.

Next, we used a combination of spectral type and temperature relations to transform the observed absolute magnitudes by \cite{HAW:02} in $r^\star i^\star z^\star J$ to produce a temperature vs. absolute magnitude relation. We then used this relation to produce colors and errors for all MLT stars. Neither the LDS nor the SDSS survey provides apparent $J$ magnitudes. Future surveys however will produce near-infrared colors (y). At such a time our code can be updated and expanded.

For WD and MS stars, we use models provided by \cite{KV:06} and \cite{K:79}, respectively. These atmospheric models are grids of color as functions of temperature and gravity. We employed a 2nd/3rd order unevenly spaced 2-dim interpolation subroutine to determine both expected absolute magnitude and error in absolute magnitude given $T$ and $\log(g)$. As before, we estimated $u^\star g^\star r^\star i^\star z^\star$ colors. This allowed simple estimates of distance modulus for each pair.

Models used to compute discrete grids are available\footnote{http://www.stsci.edu/hst/observatory/cdbs/k93models.html}, derived from the Kurucz stellar atmosphere models. For MS stars temperatures range from $3,170K\leq T \leq 42,000K$ and $0\leq \log(g) \leq5$. For WD stars, two sets of grids were used: a fine resolution with range $7,000K\leq T \leq84,000K$ and  $7\leq \log(g) \leq9.5$ in $0.5$ intervals and a broad resolution set with $4,500K\leq T \leq84,000K$ and $7\leq \log(g) \leq9$ with $1.0$ gravity intervals. The fine resolution set was used for stellar temperatures with $7,000K \leq T \leq84,000K$. For temperatures above $84,000K$ and below $7,000K$, the broad resolution set was used.

\subsection{Distance}
In order to compute the distance modulus, a variety of models were tried, including a synthetic Galactic population model that includes distance as a function of mass \citep{RRD:03}. Initially, we produced a distance model that best fits the observed LDS distribution with the given parameters. The adopted distance distribution had the form:

\begin{equation}\label{eq:g1}
f(d) = \log(d)
\end{equation}

This equation was then transformed into a random variate separation generating function:

\begin{equation}\label{eq:h1}
a(X) = 10^{3.0X+1.0}
\end{equation}

Where $0 < X < 1$ and $X$ is a uniformly distributed random number. We expect this will, through random evolutionary processes, produce the synthetic lognormal distribution of distances that \cite{RRD:03} obtained. We also assume that binary and single star distance distributions are identical.

\subsection{Stellar Evolution Code}
One of the two major tasks in this project is the ability to accurately evolve a theoretical star while keeping track of its mass, luminosity, and temperature; as previous analysis used analytical, rather than physically-motivated, mass-loss rates (e.g. \citealt{DOM:63, DOB:98, RAHO:09}). For a population of binary stars this requires substantial computational resources. As mentioned above, the Single Star Evolution (SSE) program \citep{HPT:00} was adapted for this effort. The basis of the SSE is a set of comprehensive analytical formulae, that accurately track the evolution of a star with given initial mass and metallicity. The goal of the SSE program was to allow for the rapid calculation of a number of stellar properties for a given stellar lifetime, thus facilitating the production of a population synthesis model that could produce a statistically significant sample of binaries in a reasonable computational timeframe.

The key to the SSE code's accuracy and speed is its process style, which employs a combination of interpolation formulae and tabular construction. The HPT group uses tabulated stellar models \citep{SSMM:92, CMMSS:93, MSM:98}. \cite{PSH:98} later expanded the range of mass and metallicity of the tables incorporated in the SSE program. Analytic approximations to the movement of a star in the Hertzsprung-Russell diagram (HRD) are a simplistic approach to a complex problem \citep{EFT:89}. However, supporting these calculations with tabulated stellar models allows a more precise estimation of physical properties and greatly decreases computational time.

The SSE does not take into account stars with solar masses lower then $0.2M_\odot$. For stars with masses lower then this limit we adopted the \cite{BCAH:98} low-mass stellar evolutionary models. We do not expect such low mass stars to undergo stellar mass evolution, but we do expect small changes in other stellar parameters over time. Using these evolutionary models we were able to determine the initial and final values of temperature and gravity based on initial mass and metallicity.

We selected the following parameters for the SSE code: no wind-loss, a Reimers Coefficient for mass-loss of 0.5, and a constant metallicity distribution of 0.1. This set of parameters provides an initialization point from which to develop further analysis, while still generating realistic results. Future research may include analysis (ANOVA) to determine of optimal parameter distributions and values for comparison of theoretical and observed population distributions. The authors of the SSE code have subsequently developed a follow-up algorithm, the Binary Star Evolutionary code. For development purposes, this new program was not used, as the primary concern of the DS-code is the generation of evolved binary populations and information, relating to fragile coeval binaries. The authors feel that wrapping the elliptical orbital computations, stellar population generating functions, and apparent magnitude translations around the original SSE code, allows for the maximum amount of computational efficiency and provides the necessary forensic test points for further experimentation, specifically in the generation of an ANOVA experiment for optimal parameter and distribution selection.

\subsection{Orbital Evolution}
In the present model, we assumed isotropic non-conservative mass-loss for both stars, both during the MS and PMS lifetimes. Since we modeled what could be considered two phases of mass-loss, slow (MS) and sudden (PMS), we found a two stage evolutionary code was needed.  Following \cite{VG:88, HAD:66, KOP:78}, we considered the effect double, non-interacting, mass-loss, has on the orbital parameters of the binary. For this experiment we modified versions of \cite{VG:88} equations into the simulation code presented here. We allowed no accretion onto either star from its companion and no accretion from the ISM or any outside bodies. The addition of perturbation forces from exterior sources beyond those addressed within this work, such as loose tertiary components, dense stellar populations, or other “chance” encounters were not modeled here, however we would recommend \cite{HEG:96} as starting research in the matter.

We can show via adjustment of the mass-loss orbital parameter equations that the perturbing force due to an isotropic variation of mass is tangential to the orbit, thus only the planar orbital elements will be perturbed. This assumption holds when the mass-loss timescale of the binary is much greater than the orbital timescale (period). These perturbation equations are given as a function of time, the logarithm of the total system mass, the eccentric anomaly, and the eccentricity are given as the equations \ref{eq:i1} and \ref{eq:j1}.

\begin{equation}\label{eq:i1}
\frac{d\log{a}}{dt} = -\frac{1+e\cos{E}}{1-e\cos{E}}\frac{d\log{M}}{dt}
\end{equation}

\begin{equation}\label{eq:j1}
\frac{de}{dt} = -\frac{(1-e^2)*\cos{E}}{1-e\cos{E}}*\frac{d\log{M}}{dt}
\end{equation}

Where $\frac{d\log{M}}{dt}$ is the rate of mass-loss over the time period of interest (iteration), and $E$ is the eccentric anomaly. Note, that when expressing these equations in terms of the eccentric anomaly, Kepler's equations is no longer valid and should be replaced with the equation, 

\begin{equation}\label{eq:k1}
\frac{dE}{dt} = \left( \frac{2\pi}{P} + \frac{d\log{M}}{dt}\frac{\sin{E}}{e}\right)*\left(1-e*\cos{E}\right)^{-1}
\end{equation}

The effect that this MS mass-loss has on the binary system is documented by \cite{VER:69}, \cite{DOM:63}, \cite{DOB:98} and \cite{RAHO:09}. Here, Verhulst focused the majority of the research on the effect a Jeans-Eddington mass-loss relation would analytically have on the eccentricity of the binary. It was shown that one can expect periodic variation about a mean eccentricity for a binary with isotropic continuous mass-loss depending on the parameters of the Jeans-Eddington mass-loss relationship. \cite{HAD:63} demonstrated similar effects, showing that with decreasing system mass the eccentricity of the system will sinusoidally increase. More recently \cite{VEA:11} demonstrated orbital parameter evolution for low q-ratio binaries ($q < 0.1$, exoplanets and smaller bodies). They showed that, at least for these conditions, the effect of mass-loss, varies greatly depending on the initial eccentricity distribution. 

Given that our mass-loss is a function of the stellar evolution code designed into the SSE, with the additional consideration of the low mass evolution models provided by \citep{BCAH:98}, we expect a similar result (increasing non-linear, sinusoidal eccentricity as a function of mass-loss). However we expect that the numerical (and probabilistic) nature of our experiment would demonstrate a different (and potentially more complete) relationship. For situations where mass-loss occurs on a timescale less than that of the period of the binary, we follow the “sudden” mass-loss equations \ref{eq:l1} and \ref{eq:m1} outlined by \cite{HIL:83}.

\begin{equation}\label{eq:l1}
a = \frac{a_0}{2} \left(     \frac{1 - \frac{\Delta M}{M}}{\frac{1}{2} - (\frac{a_0}{r} ) *\frac{\Delta M}{M}}     \right)
\end{equation}

\begin{equation}\label{eq:m1}
e = \left(1 - (1-e_0^2)*\left[\frac{1 - \frac{2a_0}{r}\frac{\Delta M}{M}}{(1 - \frac{\Delta M}{M})^2}\right]\right)^{0.5}
\end{equation}

We have further transformed the original equations into a function of the mass-loss ratio over a given defined time period (delta mass over original mass). The increase in separation and eccentricity, is a function of how close to periapsis or apoapsis the binary is at the time of mass-loss. It can be shown that for binary separation distances less than the semi-major axis (which is dependent on the eccentricity); sudden mass-loss can drive the binary together (decrease orbital separation). 
Similarly, as stated by \citeauthor{HIL:83}, if mass-loss occurs in a binary with a very eccentric orbit, the average post-explosion eccentricity will be less than the pre-explosion eccentricity. Conversely, given the position of the binary, the separation can be greatly amplified, resulting in populations of $WD+MS$ and $WD+WD$ pairs with separations on average larger than their $MS+MS$ counterparts. We can demonstrate for given eccentricity and location in the orbit, post-main sequence mass-loss can result in a perturbation of the orbital eccentricity beyond a bound elliptical orbit, i.e., $e \geq 1$. This is demonstrated in Equation \ref{eq:n1}.

\begin{equation}\label{eq:n1}
\frac{r}{2a} < \frac{\Delta M}{M}
\end{equation}

It is apparent, that increasing eccentricity can pose a problem for the binary system. The potential for binary separation can result from increasing the eccentricity of the binary beyond $(e \geq 1)$. Thus, we expect that the increase in eccentricity, and therefore the associated orbital velocities, in combination with the wide orbital amplitudes, will result in a number of binary systems being perturbed out of their orbits (dissolution via external perturbation). It should be noted that decreases in eccentricity resulting from sudden PMS mass-loss, coupled with the lack of mass-loss associated with white dwarf components, can result in the effective circularization of the orbit. This is apparent from the disproportionate number of binaries in the $WD+MS$ and $WD+WD$ population evolving to a final eccentricity under 0.1 in our simulations.

To model the effect high eccentricity has on the binary system, specifically on  the dissolution rate of the binary populations resulting from isotropic mass-loss and galactic tidal forces, we have implemented a function that flags evolved binary systems which have eccentricity greater than 0.95. This, combined with the \cite{AL:72} tidal limit provides a first order approximation to dissolution effects on the binary systems. Other improvements are outlined in the Binary Dissolution section

\subsection{Binary Dissolution}
Even a simple model for the orbital evolution of wide binaries must include the influences of stellar encounters, Giant Molecular Clouds (GMCs) and the Galactic tidal field \citep{WEIN:88}. For the wide separations of FBs, even minor forces such as these operating over the course of millions of years, can affect the binary separations, eventually dissolving them. \cite{AL:72} described a tidal limit for wide binaries, where the gravitational forces from Galactic tidal perturbations outweigh the gravitational attraction between the binaries themselves(Equation \ref{eq:o1}). 

\begin{equation}\label{eq:o1}
a_T \sim 2.07 \times 10^5 A.U. \left( \frac{M_1+M_2}{M_\odot}\right)^{1/3}
\end{equation}

This relation does not include either GMC or stellar encounters, but it does provide an initial cut off for galactic tidal disruption as a function of mass. We take the \cite{AL:72} equation as a first approximation to a dissolution algorithm. A more realistic approach would include all disruption effects. A rigorous treatment is beyond the scope of the current project. For a first-order approximation to the dissolution time problem, we used the \cite{WW:87} Galactic tidal perturbation formula for a $2M_\odot$ binary system with $q = 1$. The expected half-life ($t_{1/2}$) of an FB system with no consideration for the GMC encounter is given by Equation \ref{eq:p1}.

\begin{equation}\label{eq:p1}
t_{1/2} = 4.3 \times 10^3 Myr \left(\frac{a}{2.07\times10^4}\right)^{-1.4}
\end{equation}

For wide binaries with consideration for the GMC local density of $3.6\times10^8 pc^{-3}$, this relation becomes Equation \ref{eq:q1}.

\begin{equation}\label{eq:q1}
t_{1/2} = 1.32 \times 10^3 Myr \left(\frac{a}{2.07\times10^4}\right)^{-1.4}
\end{equation}

We define $t_{1/2}$ as the $50\%$ likelihood survival time for a binary with separation $a$ in $A.U.$ While this cannot be used to estimate the expansion caused by a GMC and perturbations, using a binomial distribution selection program we produced a dissolution algorithm to remove binaries from the sample that likely would have dissolved in the evolutionary time frame given. An evenly distributed binomial random variable with two possible outcomes ($50/50$) was used to simulate $t_{1/2}$.

Lastly we chose a simple ratio of survival times to linearly scale the probability mechanism. Binaries with a separation greater than the $50\%$ likelihood survival time will have a ratio greater than one and a better chance of dissociating (i.e. $50/50$ turns into $75/25$), while binaries with a separation less than the half point limit, will have a ratio less than one and a worse chance of dissociating (i.e. $50/50$ turns into $25/75$). For example: if a binary has a separation of $10^4 A.U.$ its half-life is $3655 Myr$. If a binary evolves for $3655 Myr$, there will be a $50\%$ chance of it being selected to dissolve. If the binary evolves for $10,000 Myrs$ then: $\frac{3655}{10000} = 0.37$ and $50\ast0.37\sim18$. Therefore instead of a  $50\%$ chance of our binary surviving, there is an $18\%$ chance of it surviving. Binaries that have been selected using our binomal random number generator, are removed in the posterior population samples. 

We note again that the dissolution algorithm is not a continuous function; its purpose is to generate in the final sample the expected number of dissolved binaries in a given population. The linear estimation of what should be an exponential (Poisson) function, will cause the dissolution algorithm to under-sample the number of dissolved binaries in our distribution. The \textit{dissolve} or \textit{survive} approach also does not allow for external forces and perturbations to expand the binaries. However, as stated above, the residual term from our model vs. observation comparison will provide some insight to the degree of which the sum of these external forces will effect our model. Binaries that have been removed are not considered in the following analysis.

\section{Results}
Rather then modeling individual binaries, we attempted to produce theoretical separation frequency distributions to compare with observational statistics. Histogram bin sizes were chosen to be similar for like quantities across the various cases. We then transformed our model parameters into model observations and compared our results with the observed \cite{LDS:69} separation distribution. The statistical characteristics of the distribution of primary masses, secondary masses, separations and periods for the evolved binaries is given in Table \ref{tab:b}.

\subsection{Mass Graphs}
For comparison purposes, we plotted orbital frequency distributions for both the initial and posterior populations. Figure \ref{fig:a} and Figure \ref{fig:b} include not only the initial burst binaries but also those born in bursts later on. Thus, not all binaries formed at $t_0 = 10^4$ in the initial sample, while the posterior populations have identical $t_{final}$. 

Figure \ref{fig:a} and Figure \ref{fig:b} are a representation of the same sample at an age of $\sim10 Gyr$. In Figure \ref{fig:b}, the random q-ratio has produced, in the secondaries, a distribution similar to the primaries. There are two artifacts in our secondary mass population that may not be physically realistic. First is the cutoff at $0.07M_\odot$, resulting from adopting the limit suggested by \cite{CPH:03}. However, two points mitigate this arbitrary cutoff: (1) as shown by  \citeauthor{CPH:03}, the mechanism of star formation does have a minimum mass, and (2) imposing our q values to be a number between 0 and 1 causes the average secondary mass value to be less than the primary, as expected. According to  \citeauthor{CPH:03}, the initial mass function at lower masses is poorly determined. Moreover, the LDS and SDSS surveys have detected relatively few M, L, and T dwarfs, particularly in binary systems, and not yet enough to warrant an attempt to model this region.
  
A look at the posterior distributions in Figures \ref{fig:a} and \ref{fig:b} shows the impact of mass loss. Peaks in the distribution near the $0.5M_{\odot}$ to $1.0M_{\odot}$  range and a decrease of high mass stars is a sign that some of the MS stars have evolved to less massive WD stars, or Neutron Stars and Black Holes (NS and BH). At $10 Gyr$, in some pairs both components have evolved to $\sim0.6M_{\odot}$ ($WD+WD$).

\begin{table}[h!]
\caption{Statistical Characteristics of Evolved Distributions.}\label{tab:b}
\begin{center}
\begin{tabular}{c|cccccc}
\multicolumn{7}{c}{}\\
\hline
Metric & Average & StdDev & Skew & Kurtosis & Min & Max \\
\hline
Mass 1 $(\log{M_{\odot}})$&	-0.36&	0.30&	-0.40&	-0.36&	-1.10&	1.01\\
Mass 2 $(\log{M_{\odot}})$ &	-0.58&	0.33&	0.27&	10.990&	-1.10&	0.81\\
Separation $\log{A.U.}$&	2.97&	0.55&	0.069&	-1.04&	1.74&	4.88\\
Period $\log{Myr}$ &	-1.48&	0.84&	0.075&	-1.02&	-3.60&	1.11\\
\hline
\end{tabular}
\end{center}
\end{table}

\subsection{Mass Ratio}
Figure \ref{fig:c} is a histogram of the mass ratio distribution for the initial population of binaries and the posterior population of binaries. There is a trend towards $q\sim1$ in the initial population, as expected from the mass ratio generating function. A unique by-product of evolution is evident in the posterior population of the $q$-distribution: high mass MS stars have evolved into lower mass WDs. Now the secondary has a higher mass than the WD, resulting in a distribution of binaries with $q > 1$. This time, however, the  $q > 1$ range represents the region where the primary has undergone PMS mass-loss and now has less mass than the secondary, i.e. one or both of the components of the binary have evolved into degenerate stars (WD, NS, or BH).  The additional evolution to WD of the secondary would likely only cause a density change in the population above $q=1$ or in the compact region near $q = 0.9$.

\subsection{Separation and Period}

Figures \ref{fig:d} and \ref{fig:e} present comparisons of the initial and posterior populations of binaries, the separations (in $\log{(A.U.)}$) and the period distributions (in $\log{(Myrs)}$). The sharp observed initial distribution of separations is a result of the initial parameters: a cutoff at $10^4 A.U.$ (GMC birth cloud), and a lower cutoff ($100 A.U.$) based on the definition of what qualifies as a FB. We impose an \"{O}pik distribution on top of this to produce the desired frequency distribution. The scatter is a result of binning and low number density. At higher densities ($10^5$ binaries and greater), this scatter becomes negligible. Our initial distributions are leptokurtic (i.e. flat-top), which is expected in $\log{(a)}$ due to our initial assumptions of a uniform log distribution. The skewness of the initial distribution $\sim0$ results from the flat sides and little or no discernable tail of the distribution. We see similar characteristics in the initial $\log{(P)}$ distributions(Figure \ref{fig:e}), however the edges of the distribution have been rounded out resulting from the effect the mass distribution has on the generated distribution. Our initial limits in the separation distribution results in the min ($a > 10^2 A.U.$) and max ($a < 10^4 A.U.$), respectively.

Comparisons to posterior distributions in  Figure \ref{fig:d} and Figure \ref{fig:e} demonstrate that the maximum separation is now $\log{(A.U.)} = 4.88$ ($75,850 A.U.$). Two other parameters have also changed: the shift to higher separations has depleted the number of close binaries and increased the higher separation binaries. This has effectively rounded out the histogram of separtions (kurtosis $= -1.04$) and narrowed the histogram, (standard deviation  $= 0.55$) compared to the standard deviation of $0.58$ in the initial distribution of Figure \ref{fig:d}. Figure \ref{fig:e} shows a similar trend in variance, average and kurtosis. The period has the addition of a randomly generated mass factored into its calculation; this in turn has affected the kurtosis and skewness which now reflect a more normal distribution,  per the central limiting theorem (CLT), as well as making the curve appear smoother.

\subsection{Analysis by Binary Category}
\begin{figure}[h!]
  \centering
    \includegraphics[scale=1.0]{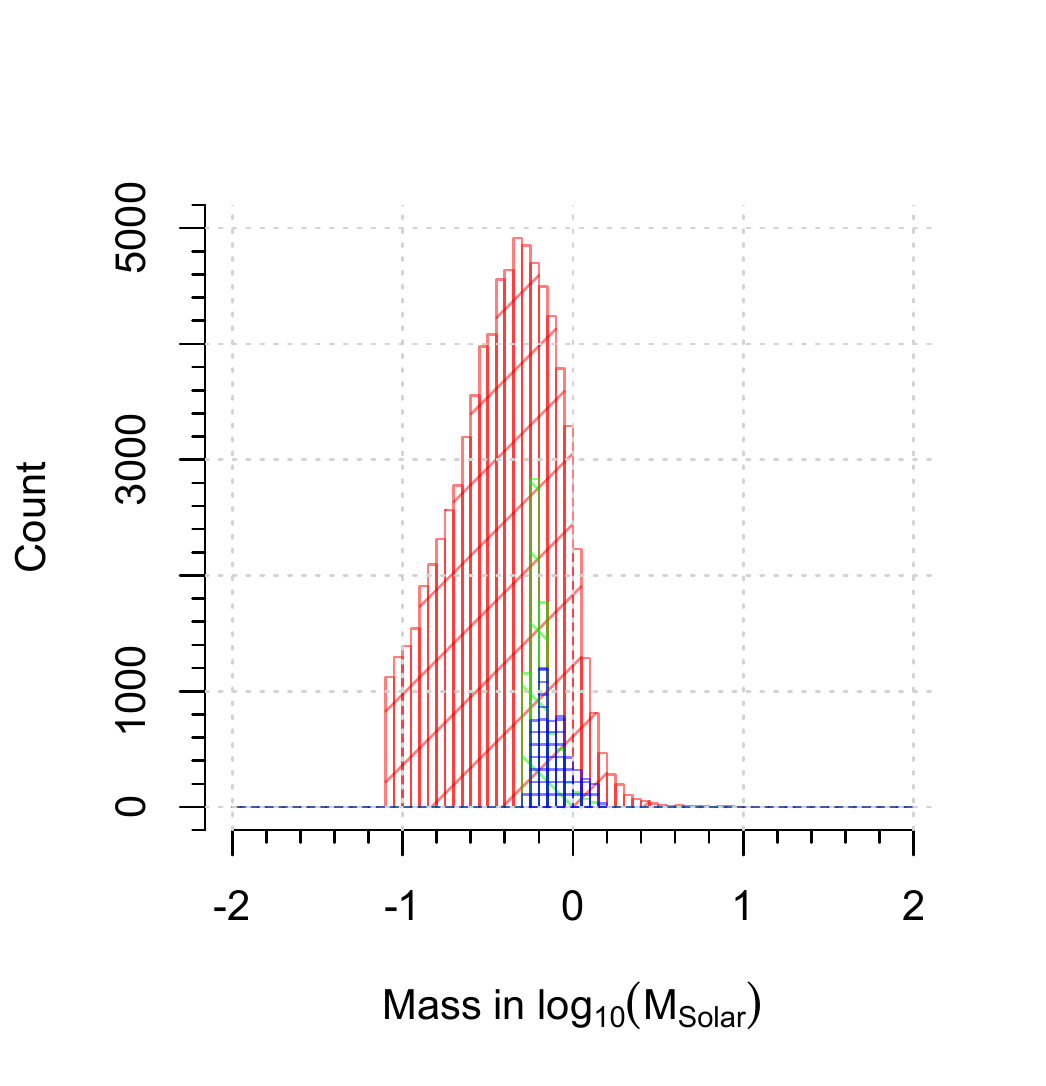}
 \caption{Histogram of the distribution of primary masses from the MS+MS populaton (in red with 45 deg. hatching), from the WD+MS population (in green with 135 deg. hatching) and the WD+WD population (in blue with 0 deg. hatching). Note the mixed colors represent overlaping distributions, bin sizes are in 0.05 $\log{(M_{\odot})}$ intervals.}\label{fig:f}
\end{figure}

\begin{figure}[h!]
  \centering
    \includegraphics[scale=1.0]{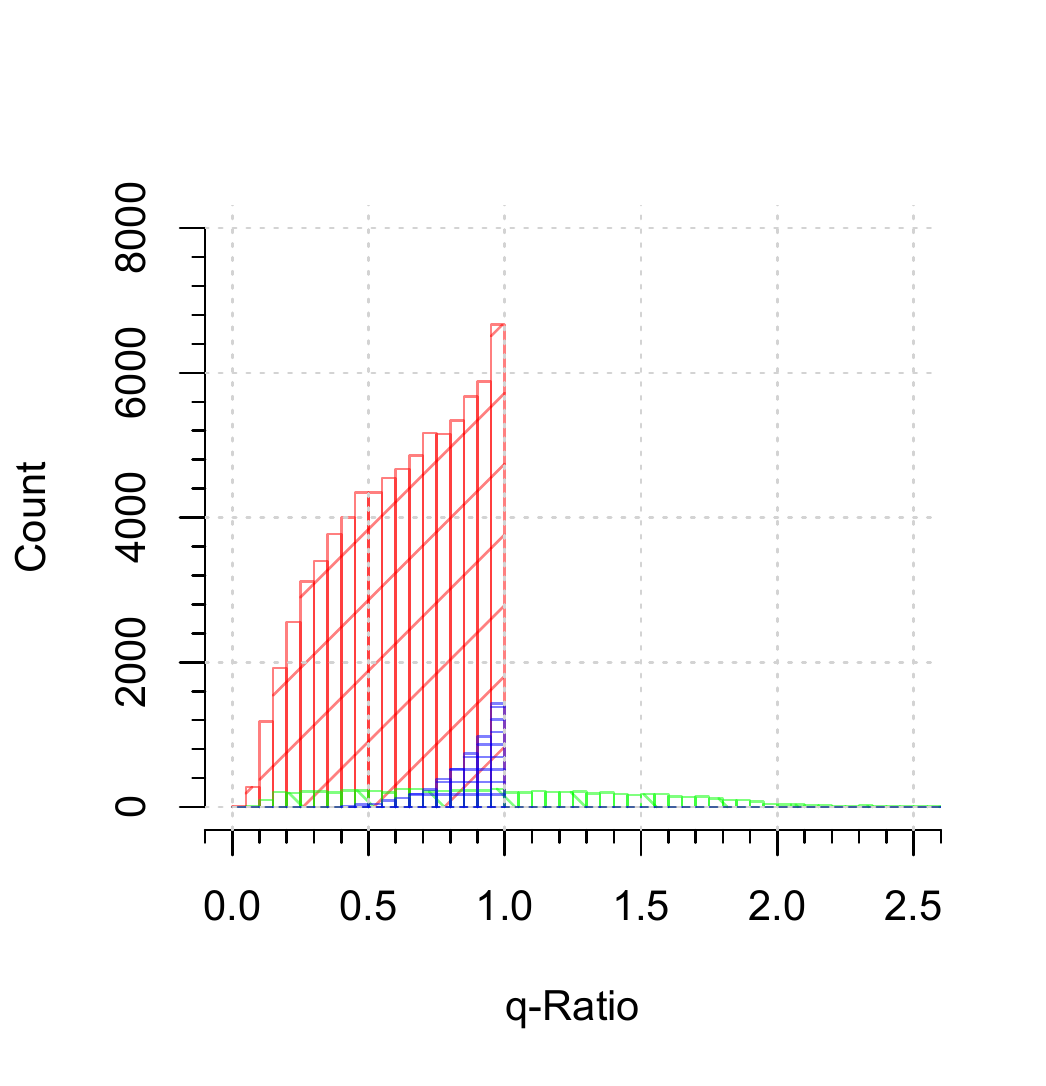}
 \caption{Histogram of the distribution of mass ratios from the MS+MS populaton (in red with 45 deg. hatching), from the WD+MS population (in green with 135 deg. hatching) and the WD+WD population (in blue with 0 deg. hatching). Note the mixed colors represent overlaping distributions, bin sizes are in 0.05 intervals.}\label{fig:g}
\end{figure}

\begin{figure}[h!]
  \centering
    \includegraphics[scale=1.0]{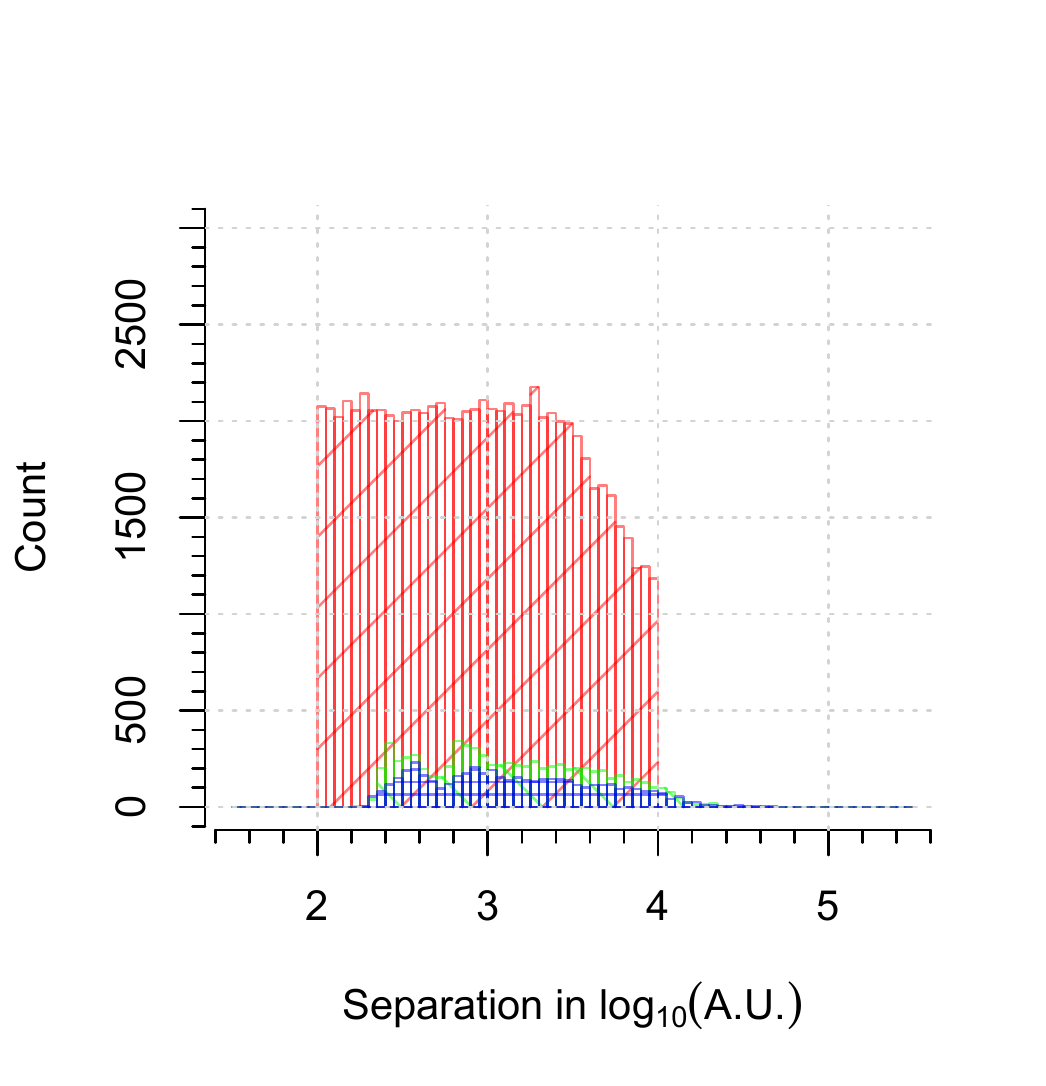}
 \caption{Histogram of the distribution of separation from the MS+MS populaton (in red with 45 deg. hatching), from the WD+MS population (in green with 135 deg. hatching) and the WD+WD population (in blue with 0 deg. hatching). Note the mixed colors represent overlaping distributions, bin sizes are in 0.05 $\log{(A.U.)}$ intervals.}\label{fig:h}
\end{figure}

\begin{figure}[h!]
  \centering
    \includegraphics[scale=1.0]{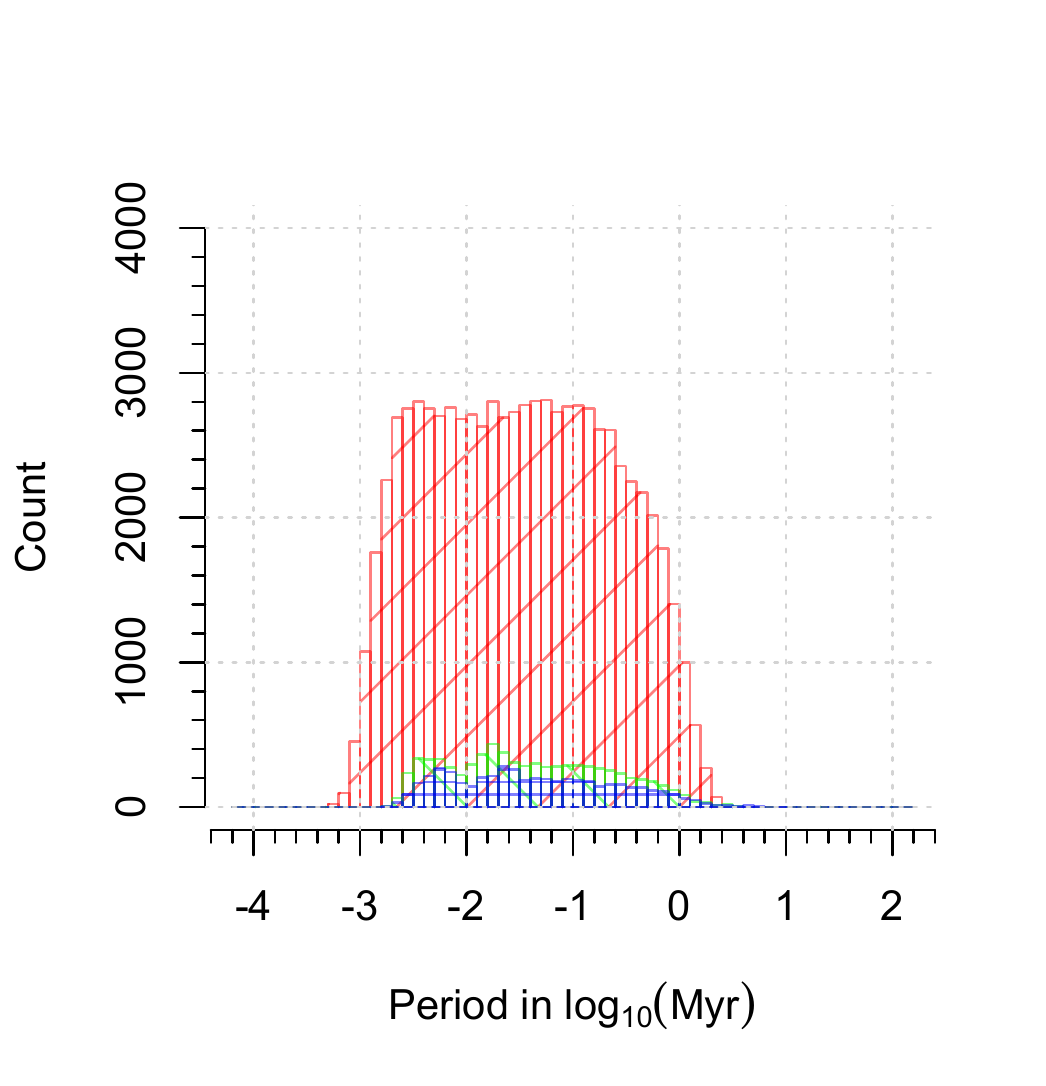}
 \caption{Histogram of the distribution of period from theMS+MS populaton (in red with 45 deg. hatching), from the WD+MS population (in green with 135 deg. hatching) and the WD+WD population (in blue with 0 deg. hatching). Note the mixed colors represent overlaping distributions, bin sizes are in 0.1 $\log{(Myr)}$ intervals.}\label{fig:i}
\end{figure}

Figures \ref{fig:f} to \ref{fig:i} are overlay plots of the three binary classes $MS+MS$, $WD+MS$ and $WD+WD$. All distrbutions are derivitives of the posterior population distribution described in the previous section; the "other" catagory refers to binaries with either component not in the MS or WD phase (RG, BH, NS...etc), the breakdown of population distribution types is given in Table \ref{tab:c}. The statistical properties of each of the three samples are given in Tables \ref{tab:d} - \ref{tab:f}.

\begin{table}[h!]
\caption{Distribution of Evolved Binaries in Posterior Population. (Population size is 90,523).}\label{tab:c}
\begin{center}
\begin{tabular}{c|cccccc}
\multicolumn{5}{c}{} \\
\hline
Metric & $MS+MS$ & $WD+MS$ & $WD+WD$ & Other  \\
\hline
Count  	&	76,886&	7,327&	4,791&	1,519\\
Percent	&	84.94\%&	8.09\%&	5.29\%&	1.68\%\\
\hline
\end{tabular}
\end{center}
\end{table}

\subsubsection{MS+MS}
The $MS+MS$ parameters are relatively unchanged and appear almost identical to the initial distribution plots in $q$-ratio, $\log{(Myr)}$ and $\log{(A.U.)}$. However, the $MS+MS$ mass plots do show missing upper mass stars. These, of course, have been transferred to the other groups as WD, BH, NS, or other stars. Similarly, a good number of higher separation binaries have dissolved, resulting in the observed change in skewness (observed asymmetry in separation distribution). Of the given model (90,523 binaries), 76,886 ($84.94\%$) binaries remain in the $MS+MS$ category. 

\begin{table}[h!]
\caption{Statistical Characteristics of Evolved MS+MS Distributions}\label{tab:d}
\begin{center}
\begin{tabular}{c|cccccc}
\multicolumn{7}{c}{}\label{tab:d} \\
\hline
Metric & Average & StdDev & Skew & Kurtosis & Min & Max \\
\hline
$\log{a} (A.U.)$ &	2.94&	0.553&	0.072&	-1.12&	2.00&	4.00\\
$\log{P} (Myr)$ &	-1.49&	0.841&	0.066&	-1.05&	-3.28&	0.39\\
\hline
\end{tabular}
\end{center}
\end{table}

\subsubsection{WD+MS}
We observe the peak mass distribution of the primary in the $WD+MS$ (i.e., the $WD$) at slightly less then unity ($WD$ with masses of less then one solar mass). In addition, the secondary is by definition an $MS$ star (has not evolved yet). The secondary mass distribution has a peak at $\sim0.7_\odot$, the approximate turn-off-mass for Disk stars, thus restricting the $q$-ratio distribution. The histogram for $WD+MS$ $q$-ratio values (Figure \ref{fig:g} ) shows a number of $q > 0$ value binaries, reflecting the fact that some original secondaries are now more massive than their evolved primaries.

\begin{table}[h!]
\caption{Statistical Characteristics of Evolved WD+MS Distributions}\label{tab:e}
\begin{center}
\begin{tabular}{c|cccccc}
\multicolumn{7}{c}{ }\\
\hline
Metric & Average & StdDev & Skew & Kurtosis & Min & Max \\
\hline
$\log{a}(A.U.)$ & 3.13&	0.489&	0.262&	-0.90&	2.24&	4.48\\
$\log{P}(Myr)$ & -1.40&	0.735&	0.246&	-0.89&	-2.78&	0.72\\
\hline
\end{tabular}
\end{center}
\end{table}

The $WD+MS$ histogram of separations shows an increase in excess kurtosis, i.e., the movement of low period and small separations towards higher values. This causes the more normal distribution curve that is seen ($kurtosis \Longrightarrow 0$). Additionally, the average separation has increased to $\log{(A.U.)} = 3.1\pm0.5$, i.e., $WD+MS$ pairs display an average separation $a = 477 A.U.$ greater than the original $MS+MS$ distribution $\log{(A.U.)}=2.9\pm0.5$. Thus, orbital amplification due to $PMS$ mass-loss is shown to have increased the average separation of the population. We note, however, that this is not the order of magnitude difference that \cite{G:86} expected. Perhaps our models have underestimated the effects of environmental perturbations (GMC, chance encounters, Galactic tidal forces). There are 7327 $WD+MS$ binaries in the final sample (i.e., $\sim8.1\%$).

\subsubsection{WD+WD}
To date, there are only about two dozen \citep{SION:91} double degenerate systems known, and certainly not as many as shown here ($N = 4791, \sim5.3\%$). The most plausible reason for this lack of discovery of $WD+WD$ binaries results from either one or both of the components being too dim to be detected in the surveys currently available. The histogram of  the $q$-ratio (Figure \ref{fig:g}) show a tendency to nearly identical primary and secondary masses. This is to be expected with the narrow range of white dwarf final masses, resulting from the SSE's IFMR calculations.

\begin{table}[h!]
\caption{Statistical Characteristics of Evolved WD+WD Distributions}\label{tab:f}
\begin{center}
\begin{tabular}{c|cccccc}
\multicolumn{7}{c}{}\\
\hline
Metric & Average & StdDev & Skew & Kurtosis & Min & Max \\
\hline
$\log{a}(A.U.)$ & 3.15&	0.500&	0.316&	-0.81&	2.27&	4.66\\
$\log{P}(Myr)$ & -1.36&	0.749&	0.308&	-0.82&	-2.74& 0.9\\
\hline
\end{tabular}
\end{center}
\end{table}

As predicted, the secondary companion (lower initial mass) produces a lower mass $WD$ than the primary, causing the peak at $q = 1$ (Figure \ref{fig:g}. Similar to the $WD+MS$ histogram of separations, the $WD+WD$ histogram shows an increase in kurtosis, i.e. loss of short period and small separations, evolving towards higher values. This causes the more normal distribution curve that is seen ($kurtosis\Longrightarrow0$) compared to the MS+MS population. The increase is the result of continued evolution of the binaries. The post-MS sudden mass-loss of the binary secondary influences the separation and period distributions a second time, pushing the uniform distribution parameters closer to a more random, Gaussian distribution. Additionally, average separation has increased from $\log{(A.U.)}=3.13$ for the $WD+MS$, to $\log{(A.U.)}=3.15$ for the $WD+WD$, i.e. an average separation increase of $63.6 A.U.$. We also note that the maximum separation has increased to $\log{(A.U.)}=4.66$ range, much greater than the initial $\log{(A.U.)}=4.0$ maximum imposed on the initial distribution.

\subsection{Comparison to Observed Data}
We begin our comparison by testing a variety of distance models against the LDS $WD+MS$ sample. For our models, we used a $10^4Myr$ evolutionary time with $10^2$ initial binaries (at $t = 0$) and a $\frac{100binaries}{10Myr}$ birth rate (i.e. constant average birth rate, \citealt{HPT:00}). For each component, we have roughly determined absolute $ugriz$ magnitudes using the \cite{K:79} models. We found temperatures and gravities from the SSE output or from \cite{CPH:03} evolutionary models for solar metallicity low-mass stars. Using the luminosities and separations, we applied a random distance model derived by assuming a uniform synthetic Solar Neighborhood population to approximate angular separation and apparent magnitude. This yielded a theoretical population histogram, restricted by the physical observation limits of the LDS. The distribution of the subset of modelled observed evolved binaries is given in Table \ref{tab:g}. Note that, in comparison to Table \ref{tab:c}, there are fewer $MS+MS$ observed than in the simulated population, an effect likely caused by the dimness of some low mass companions. Similarly, the percentage of $WD+MS$ pairs in the observed distribution is higher than in total posterior population. The latter is likely caused by the expected relative brightness of the binary pair; as mass ratio tends towards one, a pair with a $WD$ as a primary is likely to have a secondary with mass just slightly less than the progenitor of the $WD$, resulting in a brighter pair on average, than observed in the $MS+MS$ distribution.

\begin{table}[h!]
\caption{Distribution of Theoretically Observed Evolved Binarys in Postieror Population. Total observed population size is 29,6335}\label{tab:g}
\begin{center}
\begin{tabular}{c|cccccc}
\multicolumn{5}{c}{} \\
\hline
Metric & $MS+MS$ & $WD+MS$ & $WD+WD$ & Other  \\
\hline
Count &	23,507&	3,801&	1,442&	885\\
Percent&	79.3\%&	12.8\%&	4.9\%&	3.0\%\\
Fraction of Posterior & 30.6\% & 51.9\% & 30.1\% & 58.3\% \\
\hline
\end{tabular}
\end{center}
\end{table}

\subsection{Distance Model Comparison}
The comparison of our model and observed angular separation distributions is shown in Figures \ref{fig:k} and \ref{fig:l}. We expect the observational data, which is a much smaller sample ($\sim6124$ binaries) to have more random scatter than the modeled data (WD+MS population $ = 1670$ binaries). Although the agreement is good dispite the scatter in the LDS sample, there are discrepancies (that we intend to address in a later paper).

\begin{itemize}
\item The kurtosis of the model is is smaller than in observed sample. Perhaps wide binaries at smaller separations were under-observed by Luyten. Alternatively, our model's lower limit on separation could be unrealistic. 
\item The scatter in the observed data is $1\sigma$ greater than the modeled data. 
\item The mean and skewness of both data sets is similar; this suggests that our basic model is sound, at least as a first order approximation.
\end{itemize}

\section{Discussion}
The distribution of binary angular separations in our model is similar to the observed distribution(i.e., the LDS sample). The LDS sample has a lognormal separation distribution that most likely stems from a log-uniform distance distribution. The estimated range of the LDS sample is between $1<\log{d}<3$ (where distance is in parsecs), similar to our model distance range. Initially we can compare the precentage of $WD+MS$ pairs observed in the LDS sample ($511/6124 = 0.08$) with the precentage of modeled $WD+MS$ pairs ($.12$) as well as the population of $WD+WD$ pairs observed in the LDS sample ($24/6124 = 0.003$) with the population of $WD+WD$ modeled ($0.04$). These distribution numbers suggest either we have underestimated the number of binaries lost to dissolution, or over-estimated the brightiness of some of the pairs. Likewise, these comparisons may also indicate where the LDS survey is under-sampled (i.e., the WD+WD pairs).

\begin{figure}[h!]
  \centering
    \includegraphics[scale=1.0]{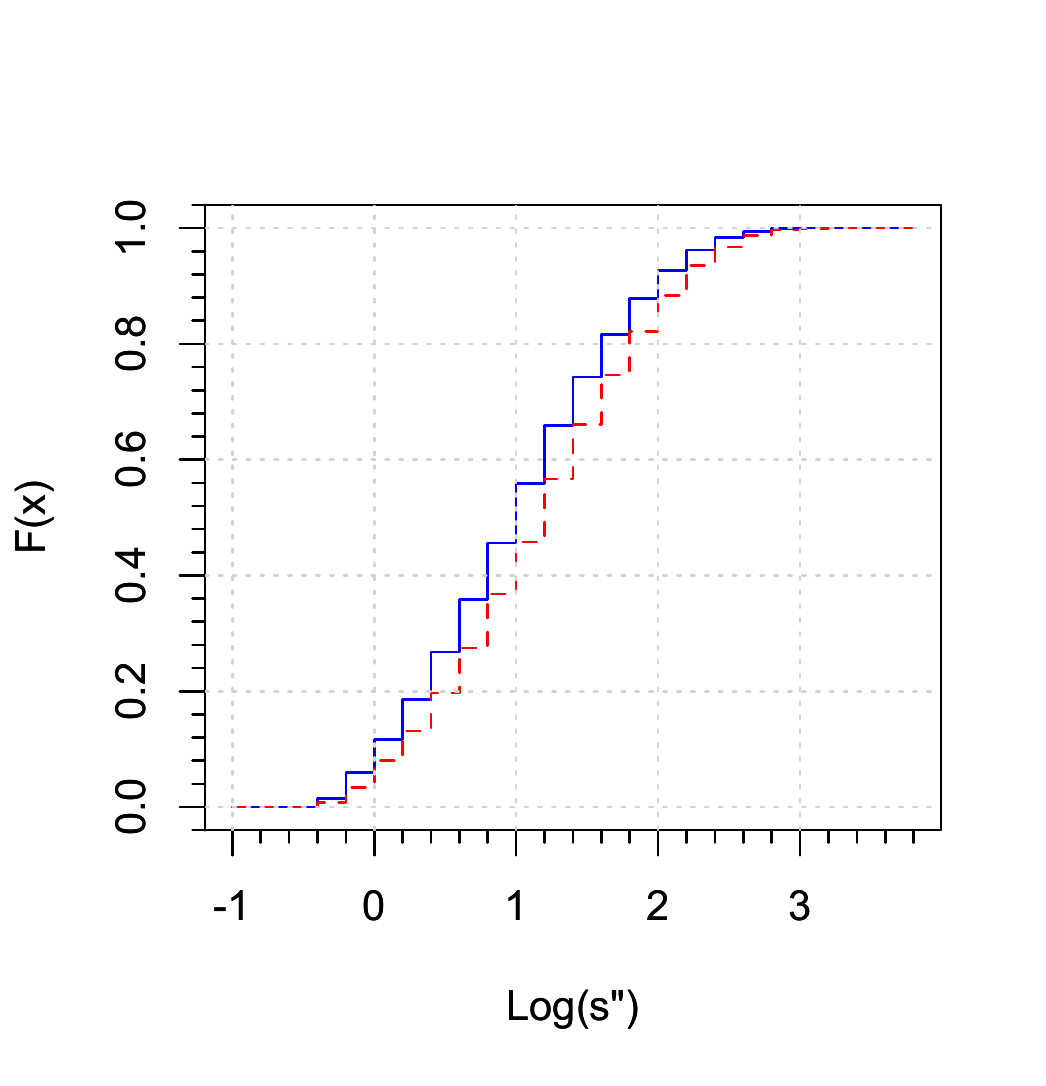}
 \caption{Comparison of observed angular separation CDF for MS+MS models and WD+MS models. Red Dashed Line: MS+MS, Blue Solid Line: WD+MS}\label{fig:j}
\end{figure}

\begin{figure}[h!]
  \centering
    \includegraphics[scale=1.0]{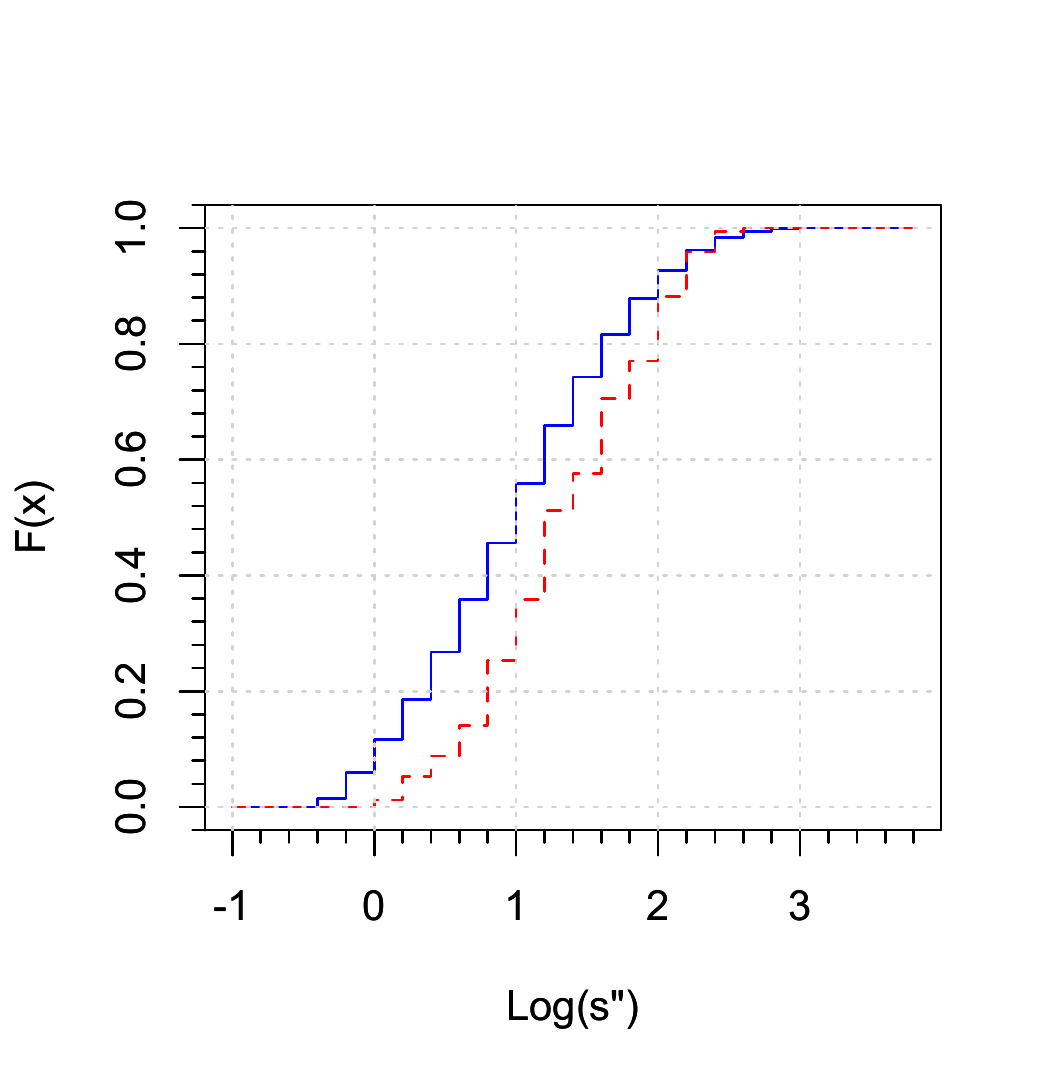}
 \caption{Plot of CDF of Model to Observed Cumulative Angular Separation Distributions for MS+MS pairs. Red Dashed Line: Observed, Blue Solid Line: Model}\label{fig:k}
\end{figure}

\begin{figure}[h!]
  \centering
    \includegraphics[scale=1.0]{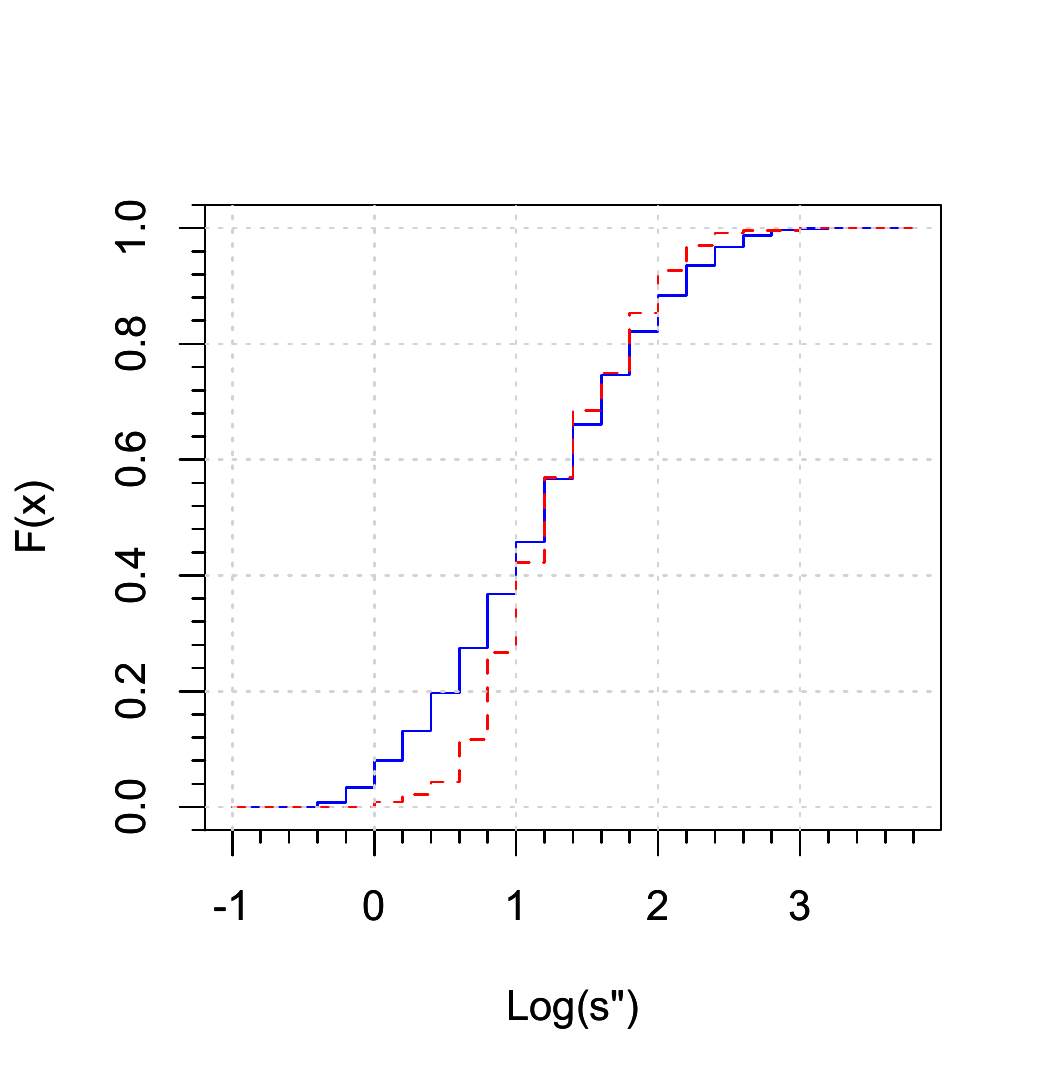}
 \caption{Plot of CDF of Model to Observed Cumulative Angular Separation Distributions for WD+MS pairs. Red Dashed Line: Observed, Blue Solid Line: Model}\label{fig:l}
\end{figure}

Figures \ref{fig:j} and \ref{fig:l}, present the results from our comparison of $MS+MS$ LDS and model pairs as well as $WD+MS$ LDS and model pairs. The figures show the cumulative distribution functions of the populations. Note that the separations are given in $\log(s")$ and have been put into 0.2 interval bins. Our initial hypothesis that PMS mass-loss significantly distorts a distribution of separations is apparent in Figure \ref{fig:k}, the offset in the CDF curves reflects an increase in average angular separation. 

We use the Cram\'{e}r-Von Mises goodness-of-fit test for comparison of the two-population cumulative distributions \citep{AN:62},  given in equations \ref{eq:r} and equation \ref{eq:s}. Here, goodness-of-fit is defined as the hypothesis $H_0: \bar{r} = 0.0$ where $\bar{r} = F_N(x) - G_M(x)$, where $F_N(x)$ and $G_M(x)$ are the cumulative distribution functions of samples of size $N$ and $M$, respectively. The Cram\'{e}r-Von Mises goodness-of-fit statistic here is defined by $T$.

\begin{equation}\label{eq:r}
T = \frac{NM}{N+M}\int_\infty^\infty [F_N(x) - G_M(x)]^2dH_{N+M}(x)
\end{equation}

\begin{equation}\label{eq:s}
T = \frac{NM}{(N+M)^2}*{\sum^N_{i=1}[F_N(x_i) - G_M(x_i)]^2 + \sum^M_{j=1}[F_N(y_j) - G_M(y_j)]^2}
\end{equation}

For the $MS+MS$ model and observation population set the values are $N = 26$ and $M = 26$. For the $WD+MS$ model and observation population set the values are also $N = 26$ and $M = 26$ as we have used the same spacing for both empirical distribution functions. These empirical distribution functions are based on the cumulative distributions generated for the populations with a bin interval of 0.2 between $0.4<\log{(s")}<2.5$.  We compute T values for the comparison of $MS+MS$ populations ($T = 0.1343$) and the comparison of $WD+MS$ populations ($T = 0.0418$). Based on these statistics we estimate the limiting distribution, $a_1(T)$, of T for our models as in \cite{AN:52}:

 \begin{equation}\label{eq:t}
a_1(T) = \lim_{n\to \infty} Pr\{n\omega^2 \leq T\}
\end{equation}

For the two comparisons, from the \cite{AN:52} tables $a_{MS+MS} > 0.5$ and $a_{WD+MS} < 0.5$, i.e., the probability of the $MS+MS$ model being drawn from the same distribution as the $MS+MS$ observed population is greater than 10\%, while the probability of the $WD+MS$ model being drawn from the same distribution as the $WD+MS$ observed population is at least 5\%. Given our original hypothesis $H_0: \bar{r} = 0.0$, the $MS+MS$ distribution comparison hypothesis fails, and the $WD+MS$ comparison hypothesis is accepted. We conclude then, that the distributions of observed and modeled $WD+MS$ angular separations are similar,  but not for the $MS+MS$ distributions.

\begin{figure}[h!]
  \centering
    \includegraphics[scale=1.0]{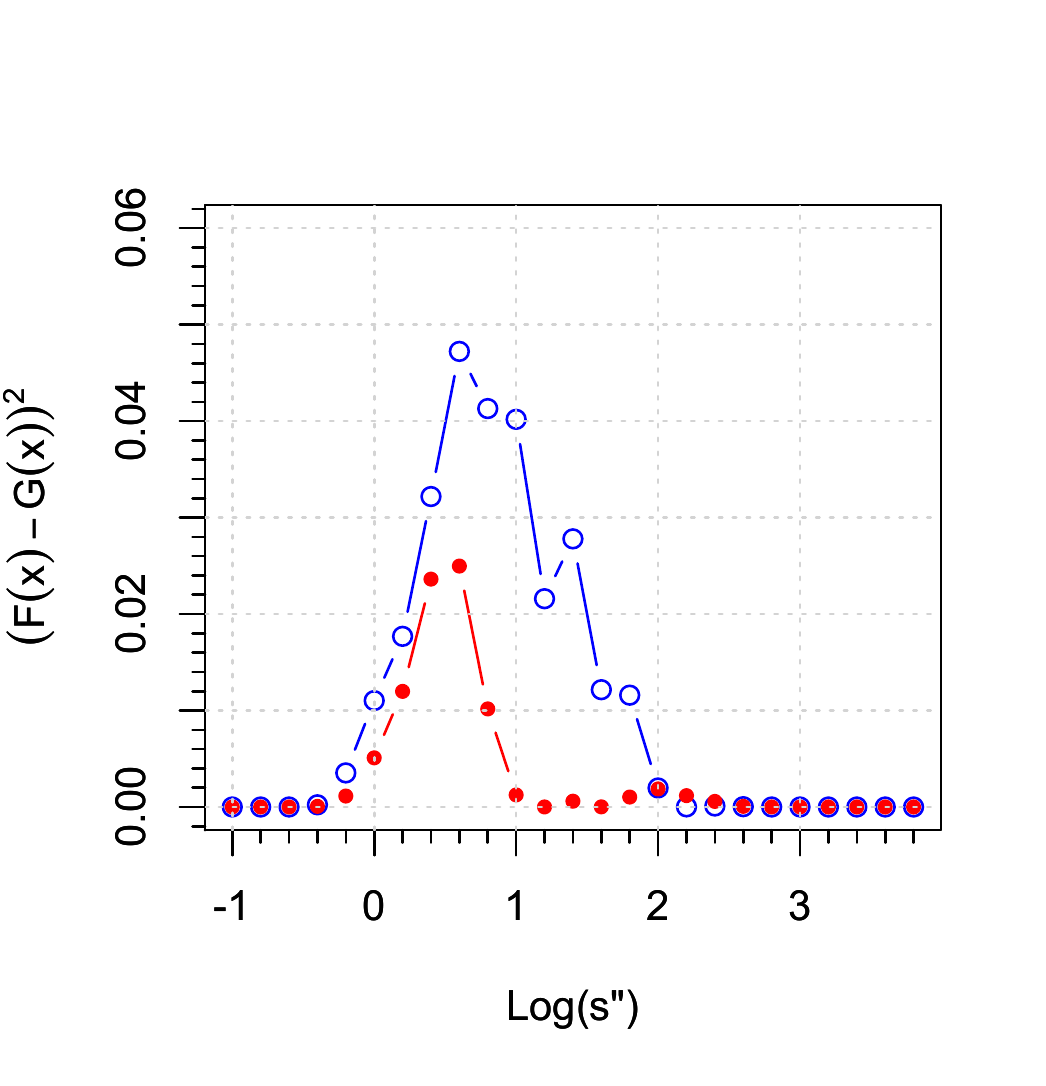}
 \caption{Plot of the Residual Square Model to Observed Cumulative Angular Separation Distributions for MS+MS pairs (in blue open circles) and WD+MS pairs (in red closed circles)}\label{fig:m}
\end{figure}

We further demonstrate the relationship between the models by graphing the residual of the angular separation distributions in Figure \ref{fig:m}. The residuals plotted have no particular linear trend, nor offset. This suggests either a difference in spread or mean. There are however, offsets in the $MS+MS$ comparison. These offsets cause the failure of the null hypothesis for the $MS+MS$ comparison.  This suggests: (1) there are additional factors for which we have not taken into consideration (such as inclination) and/or (2) small systematic errors associated with our model. 

\section{Conclusions}
We have shown that PMS mass loss significantly shifts and reshapes the distribution of $MS+MS$ separations. The semi-major axis of a wide binary is a difficult parameter to determine, requiring distance, inclination, period and other measurements. It is our hypothesis that a distribution of angular separations, preserves information about the effect of $PMS$ main-losses on the binary system. We take as proof, the following observation: Both our $MS+MS$ and $WD+MS$ models compare reasonably well to the observed LDS distributions. Our model, however, has only been affected by $PMS$ mass loss and a crude attempt has been made to account for other disrupting factors. This assumption and other first order approximation in the creation of the model, likely affect the goodness-of-fit test for the $MS+MS$ observed populations.

Figures \ref{fig:k} and \ref{fig:l} show the cumulative distribution of $\log{(s")}$ in the $MS+MS$ and $WD+MS$ pairs. The difference in observational and theoretical distributions is apparent.This difference is a function of both the incompleteness of the survey and secondary effects for which we have not taken into consideration (such as inclination or external perturbations). For example, observations and models have shown that metallicity may play a much larger role in the amount of post Main-Sequence mass-loss (see \citealt{ZH:11}). 

Future work should include a more realistic model for stellar formation history rather than the constant birth function implemented here. Furthermore, this stellar birth model should be used to drive the initial conditions of both the metallicity and the mass distribution function. In view of the trends in the metallicity, especially as a function of stellar age \citep{ZH:11}, additional effect on the mass-loss of the population as a function of time would be expected. We have demonstrated in this paper, and it has been demonstrated in other simulations, that post-MS mass-loss on average results in an increase in semi-major axis for visual and fragile binaries. Changes to the metallicity history would be expected to influence orbital amplification as a function of time, in turn affecting the final angular separation distribution. Our photometric colors were based on a consistent metallicity for the population of binaries. Adjustments to the metallicity history would affect the associated color indices, as well as the apparent magnitude of the binary components. Again, this would effect the visual angular separation distribution.

The observed $MS+MS$ and $WD+MS$ distributions are affected by observational bias and/or physical orbital amplification. If orbital amplification results in an increase in orbitla separation, we conclude that binaries with small separations have expanded to middle separation binaries, middle binary separations have increased slightly, and binaries that were already wide may have been lost to dissociation. Our current simplistic model describes the orbital frequency distribution of FBs reasonably well. In general, our biggest issue is population size of the observed distribution. The statistically small observed population causes a large scatter in the frequency distribution, or rather a larger scatter than in the model distribution. 

The DS code is currently deficient at both extremes of stellar mass ( i.e., $M < 0.07M_{\odot}$ and $M > 90M_\odot$  ). We do not have an evolutionary code applicable to these regimes, nor does our code include metallicity variation. However, such variance is not of major important in the modeling of disk stars.  Future work will require such an extension to the SSE code and extrapolation of the Kurucz color/atmosphere models. Low-mass models are available and have been taken into account for masses as low as $M\sim0.075M_{\odot}$ \citep{BCAH:98}. While their contributions to the current observed sample of FBs are small, in order to produce a realistic model of the evolved population, they should be taken into account. This lower mass limit does not directly affect the current project.

\section{Acknowledgments}
All graphs in this paper have been made with the R-project program. The authors thank Jarrod Hurley for providing them a copy of the fast stellar evolutionary code SSE. We would also like to thank Dr. Matt Wood and Dr. St\'{e}phane Vennes for their support and advice. This project was supported in part by NASA grant Y701296 (T.D.O.) and NSF grants AST 02-06115 (T.D.O.) and AST0807919.

{}

\end{document}